# Current-driven Magnetization Reversal in a Ferromagnetic Semiconductor (Ga,Mn)As/GaAs/(Ga,Mn)As Tunnel Junction

D. Chiba[1,2*], Y. Sato[1], T. Kita[2,1], F. Matsukura[1,2], and H. Ohno[1,2]

[1]*Laboratory for Nanoelectronics and Spintronics, Research Institute of Electrical Communication, Tohoku University, Katahira 2-1-1, Aoba-ku, Sendai 980-8577, Japan*

[2]*Semiconductor Spintronics Project, Exploratory Research for Advanced Technology, Japan Science and Technology Agency, Japan*

**Abstract**

Current-driven magnetization reversal in a ferromagnetic semiconductor based (Ga,Mn)As/GaAs/(Ga,Mn)As magnetic tunnel junction is demonstrated at 30 K. Magnetoresistance measurements combined with current pulse application on a rectangular 1.5×0.3 μm$^2$ device revealed that magnetization switching occurs at low critical current densities of 1.1 - 2.2 × 10$^5$ A/cm$^2$ despite the presence of spin-orbit interaction in the *p*-type semiconductor system. Possible mechanisms responsible for the effect are discussed.

**PACS:** 72.25.-b, 75.50.Pp, 85.75.-d

**Corresponding author:** Hideo Ohno

Laboratory for Nanoelectronics and Spintronics, Research Institute of Electrical Communication, Tohoku University, Katahira 2-1-1, Aoba-ku, Sendai 980-8577, Japan

tel/fax: +81-22-217-5553

e-mail:ohno@riec.tohoku.ac.jp, *e-mail:dchiba@riec.tohoku.ac.jp

Current-driven magnetization reversal is attracting great interest from the physics point of view as well as from the technological point of view. The theoretically predicted current-driven magnetization reversal by spin-transfer torque and/or spin accumulation exerted from spin polarized currents [1-3] has been observed mostly in magnetic metal current-perpendicular-to-the-plane (CPP) giant-magnetoresistance (GMR) structures [4-7]., However, the understanding of physical processes involved in the reversal has not been fully established [8]. Ferromagnetic III-V semiconductors, such as (Ga,Mn)As, are characterized by carrier-induced ferromagnetism and strong spin-orbit interaction [9], thus offering a unique combination of physics related to current-induced magnetization reversal. The ferromagnetism observed in (Ga,Mn)As is driven by the *p-d* exchange interaction and can be described by the Zener model [10], where the exchange split low energy half of the Mn *d* levels are fully occupied, which is qualitatively different from that of magnetic metals stabilized by itinerant *d*-electrons. Spin-orbit interaction has to be taken into account to describe the valence band structure of ferromagnetic III-V semiconductors, which are all *p*-type. The splitting at the top of the valence band (0.3 eV for (Ga,Mn)As) is much larger than the Fermi energy (typically 0.1 eV). The spin-orbit interaction is expected to work against current-driven magnetization reversal, because it results in mixing of the spin-up and spin-down currents, making the applicability of the successful Mott two-current approach uncertain [11, 12]. On the other hand, (Ga,Mn)As is known to have a small magnetization of 0.1 T or less [13], high carrier (hole) spin polarization $P$ [10, 14, 15], and a large *p-d* exchange interaction of the order of 1 eV, all of which act in favor of current-driven magnetization reversal. In this Letter, we present our study on current-driven magnetization reversal in fully epitaxial magnetic tunnel junctions (MTJ's) using *p*-(Ga,Mn)As.

We first describe the design of (Ga,Mn)As/GaAs/(Ga,Mn)As trilayer MTJ structure grown at 220°C by molecular beam epitaxy on $p^+$-GaAs (001) substrate. GaAs has a low barrier height of ~0.1 eV measured from the hole Fermi energy in (Ga,Mn)As [16]. These (Ga,Mn)As MTJ's exhibit high tunnel magnetoresistance (TMR)[17, 18] ratio as high as 290% at 0.4 K [15]. To optimize the thickness of the GaAs barrier layer, a series of 20 nm $Ga_{0.956}Mn_{0.044}As$ / $d$ nm GaAs / 20 nm $Ga_{0.967}Mn_{0.033}As$ MTJ's with $d$ = 1, 2, 4, 6, and 7 nm is prepared. Figure 1 shows average magnetization (normalized by its saturation value $M_S$) versus magnetic field ($M$ - $H$) curves of the samples (area ~35 mm$^2$). Inset shows the current-voltage ($I$ - $V$) characteristics of 20 × 20 μm$^2$ devices with $d$ = 2 and $d$ = 6 nm at 5 K; the qualitative features of the samples having $d$ ≤ 4 nm and $d$ ≥ 6 nm can be represented by the $d$ = 2 nm and the $d$ = 6 nm samples, respectively. The fit to Simmons' equation gives ~0.1 eV of barrier height consistent with the previous result [16]. A smooth $M$ - $H$ curve and a linear $I$ - $V$ curve are obtained for the $d$ = 2 nm sample, whereas a stepped $M$ - $H$ curve and a nonlinear $I$ - $V$ curve typical of MTJ are obtained for the $d$ = 6 nm sample. This suggests the existence of a ferromagnetic interlayer coupling when $d$ ≤ 4 nm. For the $d$ = 6 nm sample, we observed TMR ratio over 60 % at 5 K; thus $d$ = 6 nm is thick enough to decouple the two (Ga,Mn)As layers in this case. We, therefore, adopt $d$ = 6 nm in the following experiments.

For the current-driven magnetization reversal experiments, an 80 nm $Ga_{0.953}Mn_{0.047}As$ / 6 nm GaAs / 15 nm $Ga_{0.953}Mn_{0.047}As$ MTJ structure from the surface side is prepared. Different thickness for the top and the bottom (Ga,Mn)As layers is employed in order to identify the role of the total magnetic moment of the layers. The ferromagnetic transition temperatures $T_C$ of the top and the bottom layers are ~90 K and ~50 K, respectively. Rectangular devices having an in-plane aspect ratio of 5 are made with three different lateral dimensions of $a$ (// [$\bar{1}$10]) × $b$ (// [110]) = 1.5×0.3, 2.0×0.4, and 2.5×0.5 μm$^2$.

Major and minor magnetoresistance (MR) curves of a 1.5×0.3 µm$^2$ device at 30 K are shown in Fig. 2 (a) measured at a bias $V_d$ of +10 mV. Here, positive $V_d$ (and thus current) is defined as biasing the top layer positive with respect to the bottom layer. In-plane $H$ was applied along $a$ (// [$\bar{1}$10]). A square TMR curve with a TMR ratio of 15% is obtained. As shown in Fig. 2 (b), coercive force $H_C$ determined from the TMR curves increases as the device dimension reduces. By comparing the temperature dependence of $\mu_0 H_C$ ($\mu_0$: permeability of vacuum) determined from the magnetization curves with that from the TMR curves (not shown), we identify that the (Ga,Mn)As with higher $\mu_0 H_C$ is the top layer.

In our present semiconducting TMR devices, the resistance under high bias voltage is two orders of magnitude higher than the metallic structures (see the inset of Fig. 3 (a), where *I-V* characteristic of the 1.5×0.3 µm$^2$ device is shown), hence heating becomes appreciable when a dc current is used to observe the current-driven effect. Also, the strong bias voltage dependence of TMR [15, 18] makes the detection of reversal under bias difficult. Instead, we have employed the following measurement scheme: First, (1) an initial *M* configuration is prepared by applying external *H*, then (2) a 1 ms current pulse $I_{pulse}$ of varying magnitude is applied to the device at *H* = 0, and after each pulse, (3) MR as a function of *H* under a small bias $V_d$ = +10 mV with *H* // *a* // [$\bar{1}$10] is measured. Three different initial *M* configurations are employed, indicated as A, B, and C in Fig. 2 (a). Figure 3(a) shows $I_{pulse}$ dependence of Δ*R* for the 1.5×0.3 µm$^2$ device at 30 K for initial configurations A (parallel *M*, closed circles) and C (antiparallel *M*, open triangles), where Δ*R* is the difference between the resistance after application of $I_{pulse}$ and the resistance with parallel *M* at *H* = 0. A clear switching from the initial low resistance state to a high resistance state is observed in the positive current direction for configuration A. Opposite switching is observed in the negative current direction for configuration C. Note that these low and high resistances after switching correspond to those resistances of parallel and antiparallel $M$ configurations prepared by applying external $H$, respectively. In the case of configuration A (C), the initial parallel (antiparallel) $M$ alignment switches to antipallalel (parallel) at $I_{\text{pulse}} \sim +0.8$ - $+1.0$ mA ($-0.5$ - $-0.7$ mA) or at a critical current density of $J_\text{C}^\text{AP} = (+1.9 \pm 0.3) \times 10^5$ A/cm$^2$ ($J_\text{C}^\text{P} = (-1.4 \pm 0.3) \times 10^5$ A/cm$^2$). These results indicate current-driven magnetization reversal in the device. Because the reversal direction depends on the current direction, heating by current pulse does not play a major role in the observed behavior. Figure 3 (b) shows the external magnetic field dependence of critical current density $J_\text{C}$ (and critical current $I_\text{C}$) for the 2.0×0.4 µm$^2$ device. The magnetic field induced shift of $J_\text{C}$ is in the same direction as observed in the metal systems [3]. Dotted lines in Fig.3(b) are linear fits to the critical currents. The slopes (d$J_\text{C}$/d$\mu_0 H$) for $J_\text{C}^\text{P}$ and $J_\text{C}^\text{AP}$ are $\sim 3.4 \times 10^7$ A T$^{-1}$cm$^{-2}$, respectively, which are an order of magnitude smaller than those observed in metal systems.

We now take a closer look at the MR curves after application of $I_{\text{pulse}}$ greater than the threshold value. Two initial configurations (A, and B) are first prepared in the 1.5×0.3 µm$^2$ device and then a current pulse is applied. Figures 4 (a) and (b) show the results on configuration A (parallel $M$, pointing toward the positive field direction) with $J_{\text{pulse}}= +2.2 \times 10^5$ A/cm$^2$. As shown in Fig. 3, application of a positive current pulse over $J_\text{AP}$ results in antiparallel $M$ and the high resistance state. When $H$ is then applied and swept in the negative direction, the high resistance state switches to the low resistance state at −23 mT (Fig. 4 (a)), at $H_\text{C}$ of the top layer. The direction of $M$ of the top layer is thus unchanged after the application of the current pulse. When $H$ is swept in the positive direction (Fig. 4 (b)), the transition takes place at ~5 mT ($H_\text{C}$ of the bottom layer), showing that $M$ of the bottom layer was pointing to the negative direction after the current pulse. These two field sweeps demonstrate that the magnetization reversal takes place in the bottom layer, as $M$ of both layers were pointing toward the positive direction under initial configuration A. Figures 4 (c) and (d) are the results of the same measurements but starting from configuration B (parallel *M*, pointing toward the negative direction). An opposite field dependence to that in Figs. 4 (a) and (b) is obtained. This confirms again that the magnetization reversal takes place in the bottom "free" layer regardless of the initial direction of *M*. Because the current pulse direction in these experiments is fixed to positive, we can rule out the possible effect of Oersted fields generated by the pulse in the observed reversal.

Hereafter, we discuss the possible mechanisms responsible for the reversal. The effect of Oersted fields and heating by current pulse can be ruled out from the experimental observation. The most probable origin responsible for the observed reversal is spin-transfer torque exerted by spin-polarized current [1, 2]. The current direction for switching, which is the same as observed in metallic systems [3-6], is consistent with what we expect from the way the bands spin-split in (Ga,Mn)As; due to the negative *p-d* exchange interaction. The fact that the bottom thin layer is the free layer is consistent with the spin-transfer torque model as the total magnetic moment of the bottom layer is less than that of the thicker top layer.

In order to put the present results into perspective, we compare the results with the critical current densities calculated by available phenomenological formula based on spin-transfer torque applicable to metallic CPP-GMR systems. Note that no theory has been developed specifically for the ferromagnetic semiconductor system, and few theory exists for spin-transfer torque in MTJ's because this has been reported only recently even in metallic systems [19]. Because the applied bias during the switching is much larger than the barrier height in the system studied here, the threshold current for CPP-GMR systems may be a reasonable approximation for the present system. Under the assumption of coherent rotation of magnetization with a uniaxial anisotropy, the critical current densities for switching are calculated to be $J_P = -1.2 \times 10^6$ A/cm$^2$ and $J_{AP} = 3.9 \times 10^6$ A/cm$^2$ for our devices using the Slonczewski's formulae for CPP-GMR devices [1]. The parameters used are, 0.04 T for $M_S$, $P$ = 0.26 calculated from the TMR ratio of 15% using the Julliere's formula [20], and the damping constant $\alpha$ = 0.02 from the reported ferromagnetic resonance data and theories [21-23]. We employed $\mu_0 H_A = 0.1$ T as the anisotropy field [21, 24]; small shape anisotropy field (of the order of 0.01T) is neglected [25]. The calculated critical current densities are an order of magnitude smaller than those in metal systems [3-6], reflecting the small magnetization of (Ga,Mn)As, but they are almost an order of magnitude greater than the observed values, theory is clearly needed to understand the role of spin-orbit interaction in the presence of competing *p-d* exchange with it. In the present case, the situation may be more complex than the metallic cases. *P* of current pulse may even be lower than the one estimated from the TMR ratio because of the voltage dependence and contains uncertainty as *P's* are not identical for both (Ga,Mn)As electrodes. Incoherent processes during reversal might also be taking place, as the transition at the lower $H_C$ (see Fig. 2) is not as sharp as the higher one, suggesting the possibility of the nucleation of magnetic domains during switching. Competition among various magnetic in-plane anisotropies (uniaxial crystal anisotropy, cubic crystal anisotropy, and shape anisotropy) [21, 24, 25] can also be the source for the non-abrupt transition, by which trapping of local *M* along the metastable directions may occur. Local heating could affect to the reduction of threshold current, as a small temperature rise results in reduction of magnetization of the bottom free layer. The same formula gives $dJ_C/d\mu_0 H$ to be $2.8 \times 10^7$ A T$^{-1}$cm$^{-2}$ for $J_C^{AP}$ and $0.6 \times 10^7$ A T$^{-1}$cm$^{-2}$ for $J_C^P$, respectively, which is in reasonable agreement with the experimental observation. The reduced sensitivity of the critical current on the applied magnetic fields may thus be related to the small magnetization of (Ga,Mn)As.

In summary, we have shown that the current-driven magnetization reversal takes place in a ferromagnetic semiconductor magnetic tunnel junction. The most probable mechanism responsible for the observed magnetization reversal is the spin-transfer torque exerted from the spin polarized current. The critical current density for reversal is found to be of the order of $10^5$ A/cm$^2$, two orders less than the metal systems. The low critical current densities observed here may also be of technological importance, because the current-induced reversal is advantageous for ultra-high density magnetic memories over magnetization reversal using magnetic fields.

We thank M. Yamanouchi, H. Matsutera, M. Shirai, Y. Ohno, K. Ohtani, T. Dietl, M. Sawicki, J. Wróbel, and A. H. MacDonald for useful discussions. This work was partly supported by the IT-Program of Research Revolution 2002 (RR2002) from MEXT, a Grant-in-Aid from MEXT, and the 21st Century COE Program at Tohoku University.

**Figure captions**

**Fig. 1** Normalized magnetization curves of two (Ga,Mn)As/GaAs ($d$ nm)/(Ga,Mn)As (~35 mm$^2$) magnetic tunnel junction layers at 5 K ($d$ = 2 nm and $d$ = 6 nm). Magnetic field was applied along [$\bar{1}$10]. The inset shows $I$ - $V$ characteristics at 5 K of the two magnetic tunnel junction devices (20×20 μm$^2$) fabricated from these layers.

**Fig. 2** (a) Major (closed symbols) and minor (open symbols) magnetoresistance curves of a 1.5×0.3 μm$^2$ magnetic tunnel junction sample at 30 K taken at a bias of $V_d$ = +10 mV. Magnetic field is applied along $a$. Three arrows indicate the sweep direction of magnetic field to prepare the different initial configurations A, B, and C. (b) Size $A$ dependence of coercive force $\mu_0 H_C$. $H_C$ of the top (bottom) layer are indicated by closed (open) symbols.

**Fig. 3** (a) $\Delta R$ as a function of $I_{pulse}$ of the 1.5 × 0.3 μm$^2$ device at 30 K, where $\Delta R$ is the resistance difference between the resistance of MTJ after application of $I_{pulse}$ (1 ms) and that at parallel magnetization configuration at $H$ = 0. Closed circles show the $I_{pulse}$ dependence of $\Delta R$ for initial configuration A (parallel $M$), whereas open triangles show the results for initial configuration C (antiparallel $M$). The inset shows $I$-$V$ characteristics of the device. (b) Magnetic field dependence of critical current density $J_C$ (and critical current $I_C$) for the 2.0×0.4 μm$^2$ device at 30 K. Dotted lines are linear fits to the data.

**Fig. 4** Magnetoresistance curves of the 1.5×0.3 μm$^2$ device at 30 K measured at $V_d$ = +10 mV starting from three different states. MR curves (a) and (b) are obtained from a state prepared by applying a positive current pulse with current density of $J_{pulse}$ = + 2.2 × 10$^5$ A/cm$^2$ on initial configuration A (see the rightmost diagram for $M$ configuration). MR

curves (c) and (d) are obtained from a state prepared by the same manner but starting from initial configuration B.

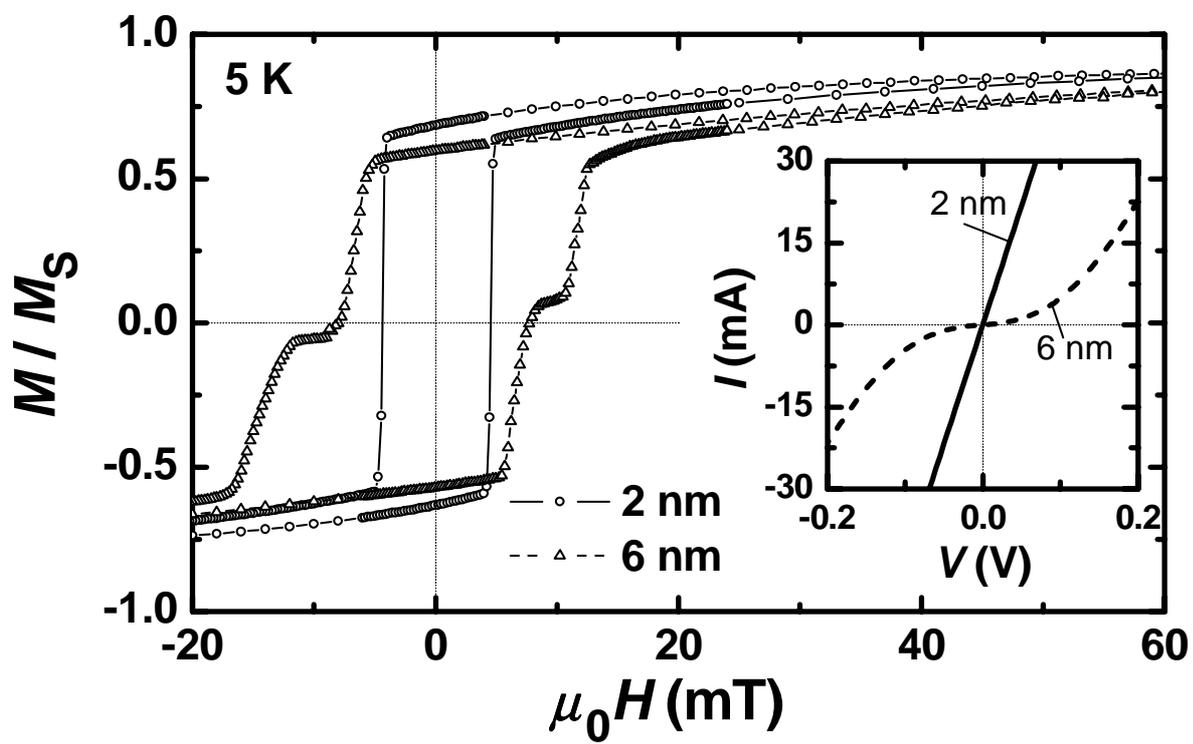

Fig.1. Chiba *et al.*

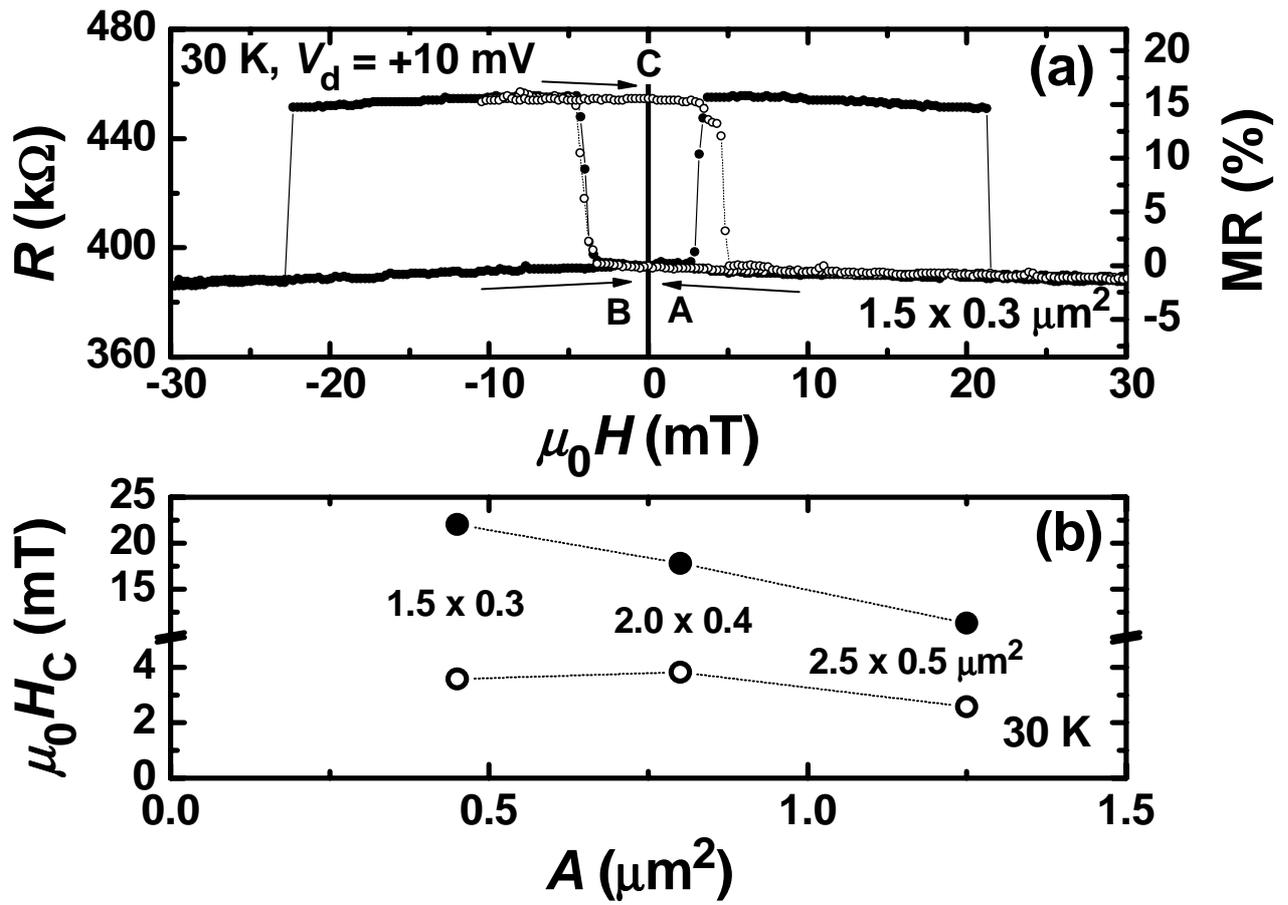

Fig.2. Chiba *et al.*

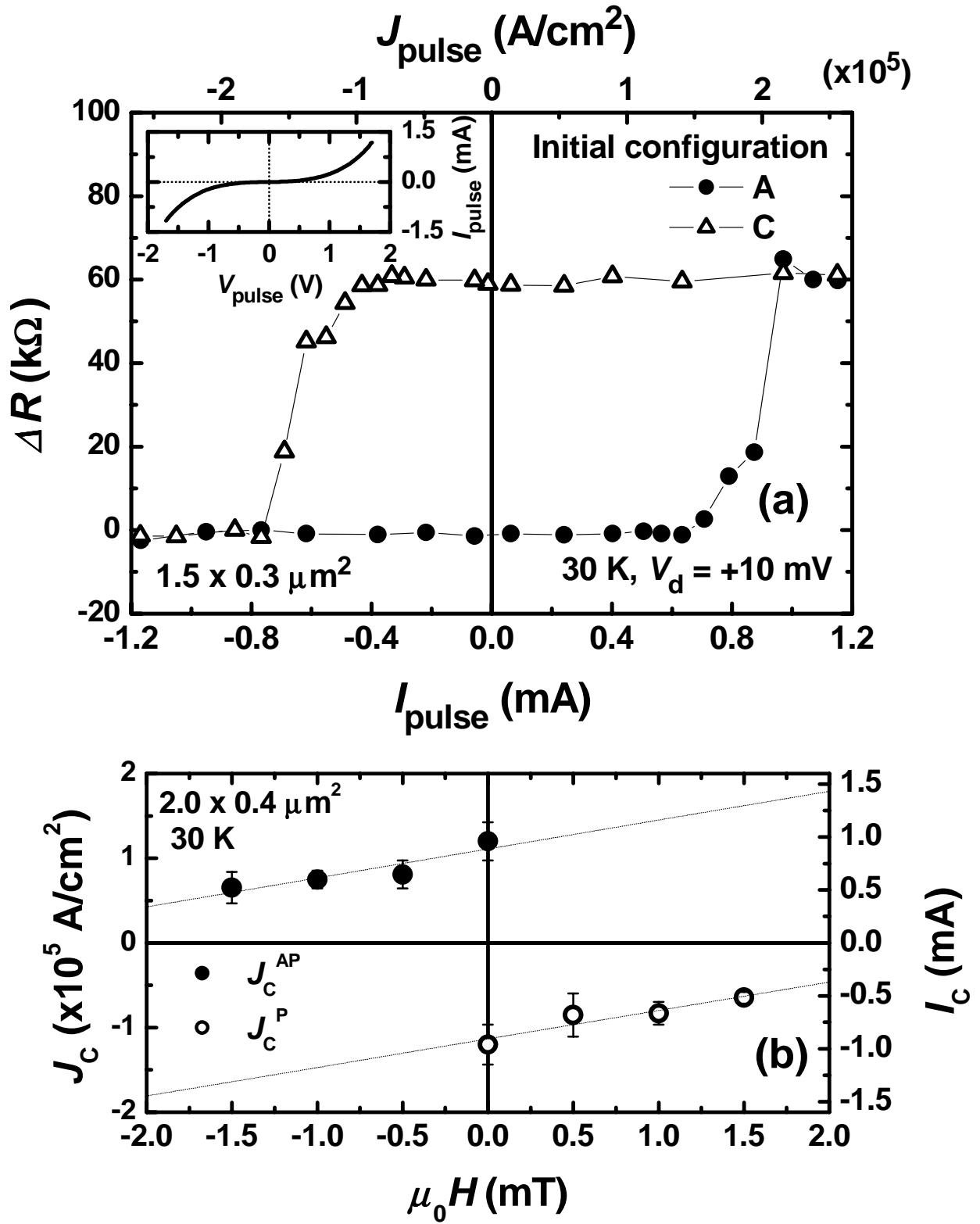

Fig.3. Chiba *et al.*

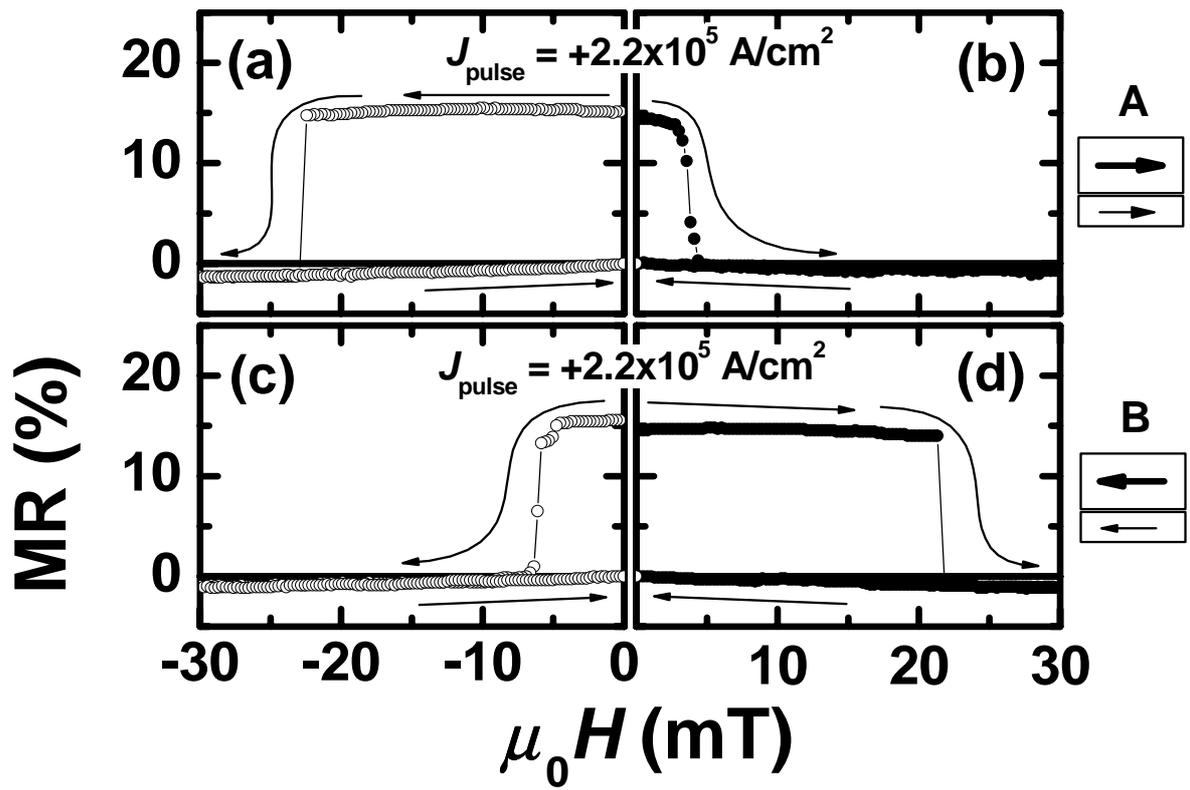